%% file: main.tex
\documentclass[twocolumn, switch]{article} % Method A for two-column formatting

\usepackage{preprint}

\usepackage{amsmath, amsthm, amssymb, amsfonts}

\usepackage[numbers,square]{natbib}
\usepackage[utf8]{inputenc}	% allow utf-8 input
\usepackage[T1]{fontenc}	% use 8-bit T1 fonts
\usepackage{xcolor}		% colors for hyperlinks
\usepackage[colorlinks = true,
            linkcolor = purple,
            urlcolor  = blue,
            citecolor = cyan,
            anchorcolor = black]{hyperref}	% Color links to references, figures, etc.
\usepackage{booktabs} 		% professional-quality tables
\usepackage{nicefrac}		% compact symbols for 1/2, etc.
\usepackage{microtype}		% microtypography
\usepackage{lineno}		% Line numbers
\usepackage{float}			% Allows for figures within multicol

\usepackage{lipsum}		%  Filler text

\usepackage{newfloat}
\DeclareFloatingEnvironment[name={Supplementary Figure}]{suppfigure}
\usepackage{sidecap}
\sidecaptionvpos{figure}{c}

\usepackage{titlesec}
\titlespacing\section{0pt}{12pt plus 3pt minus 3pt}{1pt plus 1pt minus 1pt}
\titlespacing\subsection{0pt}{10pt plus 3pt minus 3pt}{1pt plus 1pt minus 1pt}
\titlespacing\subsubsection{0pt}{8pt plus 3pt minus 3pt}{1pt plus 1pt minus 1pt}

\usepackage{tikz,xcolor,hyperref}

\definecolor{lime}{HTML}{A6CE39}
\DeclareRobustCommand{\orcidicon}{
	\begin{tikzpicture}
	\draw[lime, fill=lime] (0,0) 
	circle [radius=0.16] 
	node[white] {{\fontfamily{qag}\selectfont \tiny ID}};
	\draw[white, fill=white] (-0.0625,0.095) 
	circle [radius=0.007];
	\end{tikzpicture}
	\hspace{-2mm}
}
\foreach \x in {A, ..., Z}{\expandafter\xdef\csname orcid\x\endcsname{\noexpand\href{https://orcid.org/\csname orcidauthor\x\endcsname}
			{\noexpand\orcidicon}}
}
\title{A Toolkit for Generating Code Knowledge Graphs}

\usepackage{authblk}

\author[1]{Ibrahim Abdelaziz}
\author[1]{Julian Dolby}
\author[2]{Jamie McCusker}
\author[1]{Kavitha Srinivas}

\affil[1]{IBM Research, T.J. Watson Research Center, Yorktown Heights, NY, USA\\

\{ibrahim.abdelaziz1, kavitha.srinivas\}@ibm.com, dolby@us.ibm.com}
\affil[2]{Rensselaer Polytechnic Institute (RPI), Troy, NY, USA

mccusj2@rpi.edu
}
\input{macros.tex}

\begin{document}

\twocolumn[ % Method A for two-column formatting
  \begin{@twocolumnfalse} % Method A for two-column formatting
  
\maketitle

\begin{abstract}
Knowledge graphs have been proven extremely useful in powering diverse applications in semantic search and natural language understanding.  
In this work, we present {\logo}, a toolkit to build code knowledge graphs that can similarly power various applications such as program search, code understanding, bug detection, and code automation. {\logo} uses generic techniques to capture code semantics with the key nodes in the graph representing  classes, functions and methods. Edges indicate \textit{function usage} (e.g., how data flows through function calls, as derived from program analysis of real code), and \textit{documentation} about functions (e.g., code documentation, usage documentation, or forum discussions such as StackOverflow).  Our toolkit uses named graphs in RDF to model graphs per program, or can output graphs as JSON.  We show the scalability of the toolkit by applying it to 1.3 million Python files drawn from GitHub, 2,300 Python modules, and 47 million forum posts. This results in an integrated code graph with over 2 billion triples. We  make the toolkit to build such graphs as well as the sample extraction of the 2 billion triples graph publicly available to the community for use. 
\end{abstract}
%\keywords{First keyword \and Second keyword \and More} % (optional)
\vspace{0.35cm}

  \end{@twocolumnfalse} % Method A for two-column formatting
] % Method A for two-column formatting

%\begin{multicols}{2} % Method B for two-column formatting (doesn't play well with line numbers), comment out if using method A

%%%%%%%%%%%%%%%  Main text   %%%%%%%%%%%%%%%
% \linenumbers

\input{sections/introduction.tex}

\input{sections/ontology.tex}
\input{sections/static_analysis.tex}
\input{sections/extraction_linking}
\input{sections/stats}

\input{sections/related}

%%%%%%%%%%%% Supplementary Methods %%%%%%%%%%%%
%\footnotesize
%\section*{Methods}

%%%%%%%%%%%%% Acknowledgements %%%%%%%%%%%%%
%\footnotesize
%\section*{Acknowledgements}

%%%%%%%%%%%%%%   Bibliography   %%%%%%%%%%%%%%
\normalsize
\bibliography{paper}     

%%%%%%%%%%%%  Supplementary Figures  %%%%%%%%%%%%
%\clearpage

%%%%%%%%%%%%%%%%   End   %%%%%%%%%%%%%%%%
%\end{multicols}  % Method B for two-column formatting (doesn't play well with line numbers), comment out if using method A
\end{document}

%% file: macros.tex
\newcommand{\logo}{\textsc{GraphGen4Code}}

%% file: sections/introduction.tex
\section{Introduction}

A number of different knowledge graphs have been constructed in recent years such as DBpedia \cite{dbpedia-swj}, Wikidata \cite{Vrandecic:2014:WFC:2661061.2629489}, Freebase \cite{Bollacker08freebase:a}, YAGO \cite{Suchanek:2007:YCS:1242572.1242667} and NELL \cite{Carlson:2010:TAN:2898607.2898816}.
These knowledge graphs have provided significant advantages in a number of different application areas, such as semantic parsing~\cite{heck2013leveraging}, recommendation systems~\cite{catherine2016personalized}, information retrieval~\cite{dietz2018utilizing}, and question answering~\cite{abdelaziz2021semantic,sun2018open,wang2019improving}. 
%JPM: We don't really discuss embeddings in this paper.
%Wang \textit{et al.} \cite{journals/tkde/WangMWG17} provides a comprehensive review of the use of knowledge graph embeddings in applications.  
Inspired by the value of these knowledge graphs for a variety of  applications, we asked how one might build a knowledge graph in the domain of program code.  
There are a number of applications around code that could potentially benefit from such knowledge graphs, such as code search, code automation, refactoring, bug detection, and code optimization \cite{Allamanis:2018:SML:3236632.3212695,codebreaker}.

In 2019-2021 alone, there have been over 150 papers\footnote{\url{https://ml4code.github.io/papers.html}} using machine learning for problems that involve code, including problems that span natural language and code (e.g., summarizing code in natural language \cite{alon2018codeseq}).  Yet there are two problems with existing work on code representation: (a) they rely on local representations of code such as Abstract Syntax Trees, or lines of source text (see \cite{Allamanis:2018:SML:3236632.3212695} for a comprehensive survey on the topic). Few works have used data and control flow based representations of code as input to drive various applications although these representations capture locality in program code which is highly non-local, (b) they rarely include any natural language associated with code unless it is for a specific task such as code search ~\cite{feng2020codebert}.  As humans, we understand code by understanding documentation of individual API calls, but also its flow as objects get created and passed through program sections that are often non-local.  Our goal therefore was to build a toolkit that can construct code representations that represent actual program flow along with natural language descriptions of API calls when they exist to enhance code representations.

\begin{figure*}
\begin{center}
\centering
\includegraphics[width=0.9\linewidth]{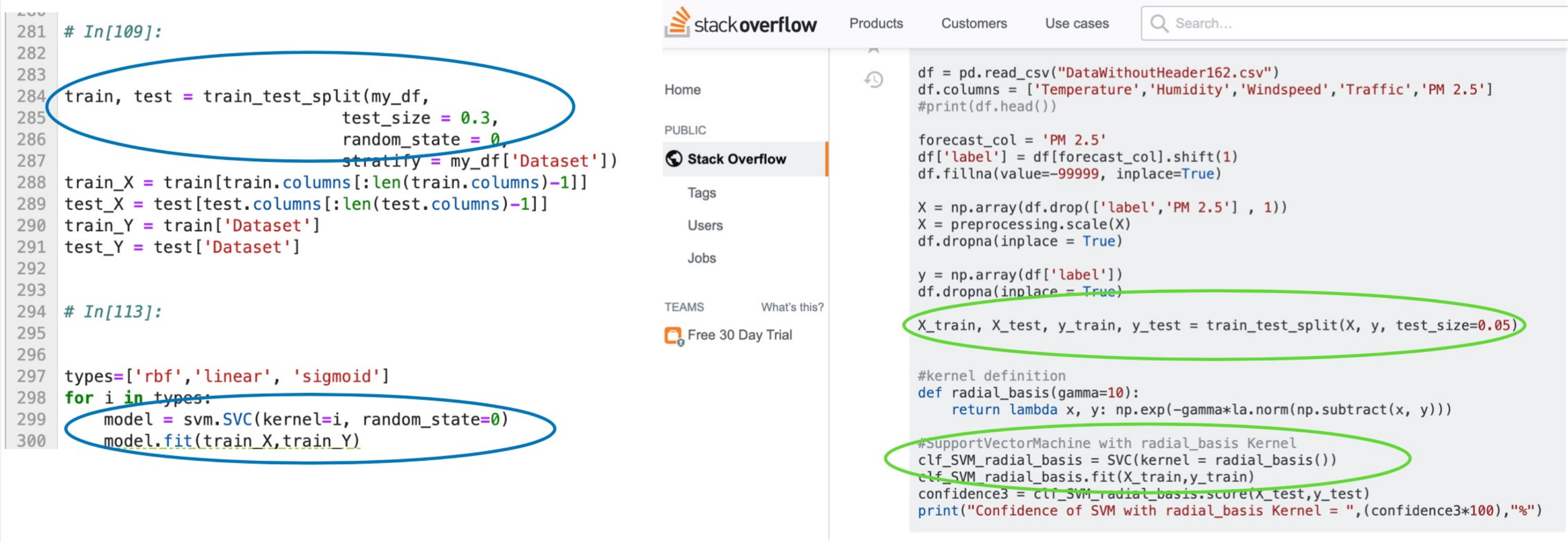}
\end{center}
\caption{Code search example: Program (left) and a relevant forum discussion (right)}
\label{fig:motivating_example}
\end{figure*}

We first illustrate the value of such a toolkit with Figure~\ref{fig:motivating_example},
which shows an example of code search: a developer searches for StackOverflow posts relevant to the Python code from GitHub in the left panel.  
That code creates an SVC model (\texttt{svm.SVC}) to train on the dataset (\texttt{model.fit}).  On the right is a real post from StackOverflow relevant to this code, in that the code performs similar operations.  However, treating program code as text or as an Abstract Syntax Tree (AST) makes this similarity extremely hard to detect.  For instance, there is no easy way to tell that \texttt{model} and \texttt{clf\_SVM\_radial\_basis} refer to the same type of object.  They look different as a token level unless one performs data flow analysis, which would have an edge from \texttt{svm.SVC} to the \texttt{svm.SVC.fit} call in both programs, and would abstract out variable names.  Furthermore, one can use natural language descriptions of \texttt{SVC} for instance, to realize it is similar to \texttt{linear\_model.SGDClassifier}, and the usage guide, or forum posts often suggest using the latter for larger datasets.
% Such an abstract representation would ease understanding of semantic similarity between programs, making searches for relevant forum posts more precise. %Yet most representations in the literature have been application-specific, and most have relied on surface forms such as tokens or ASTs.
By building a toolkit to build richer representations of code, it is possible to generate more abstract code representations than token or AST level, and these representations in turn can greatly augment code representation learning. 
% Yet, most representations used in the literature so far have been application-specific, and most have relied on surface forms of programs such as their tokens or ASTs, which greatly limit their use.

{\logo} is therefore designed as a toolkit to build knowledge graphs for program code.  We deploy state-of-the-art program analysis techniques that generalize across programming languages to build inter-procedural data and control flow.  In general, applying these techniques on a large scale to millions of API libraries is difficult because of the modeling of the semantics of individual API libraries.  In this paper, we build a set of abstractions over APIs that allow us to scale such analysis to millions of programs.  We target Python for our demonstration of scalability, because Python poses particularly difficult challenges as a dynamic language, although our techniques can be extended easily to Javascript and Java.  To demonstrate {\logo}'s scalability, we build graphs for 1.3 million Python programs (where program refers to a single Python script) on GitHub, each analyzed into its own separate graph.  We also use the toolkit to link library calls to documentation and forum discussions, by identifying the most commonly used modules in code, and trying to connect their classes, methods or functions to relevant documentation or posts.  For forum posts, we used information retrieval techniques to connect it to method or the class (an example of which is shown in Figure~\ref{fig:motivating_example} for the \texttt{sklearn.svm.SVC.fit} method).  We performed this linking for 257K classes, 5.8M methods, and 278K functions, and processed 47M posts from StackOverflow and StackExchange to show the feasibility of using the \logo toolkit for building knowledge graphs for code.  

%From public code, we extract a graph per program, which captures that program in terms of data and control flow.  As an example, the program in Figure~\ref{fig:motivating_example} has a data flow edge from a node denoting the \texttt{sklearn.train\_test\_split} to a node denoting \texttt{sklearn.svm.SVC.fit}.  It would also have a control flow edge from \texttt{sklearn.svm.SVC} to  \texttt{sklearn.svm.SVC.fit}.
%dlm the sentence above is hard to parse.  can you rewrite?  also look at usage of that/which
%Representing programs as data flow and control flow is crucial because programs that behave similarly can look arbitrarily different at a token or AST level due to syntactic structure or choices of variable names.  Conversely, programs that look similar (e.g., they invoke a call to a method called \texttt{fit}) can have entirely different meanings.  We build such representations for 1.3 million Python programs on GitHub, each analyzed into its own separate graph. 

%While these graphs capture how libraries get used, they are not sufficient; there are semantics in textual documentation of libraries, and in forum discussions of them.  
%To our knowledge, this is the first knowledge graph that captures the semantics of code, and it associated artifacts.

\sloppypar
%Our knowledge graph is an RDF graph since RDF provides good abstractions for modeling individual program data such as named graphs, and SPARQL, the query language for RDF provides extensive support for the sorts of operations that are crucial for understanding control and data flow in programs, especially transitivity.  Each program graph is modeled separately, and we use existing properties from ontologies such as schema.org, SKOS, DublinCore, and SemanticScience Integrated ontology (SIO) to model relationships between program and documentation entities.  

In summary, our key contributions are as follows:
\begin{itemize}
    \item A scalable toolkit for building knowledge graphs for code
    \item A model to represent code and its natural language artifacts (Section \ref{modeling})
    \item A language neutral approach to mine common code patterns (Section \ref{sec:analysis})
    \item A demonstration of the toolkit's scalability by applying it to 1.3 million Python programs and 47 million posts to generate a code knowledge graph with over 2 billion facts about code  (Section \ref{sec:extraction})
    %\item Use cases showing generality and promise of {\logo} (Section \ref{graph_properties})
\end{itemize}
 All artifacts associated with {\logo}, along with detailed descriptions of the modeling, use cases, query templates and sample queries for use cases are publicly available at \url{https://wala.github.io/graph4code/}.

%% file: sections/ontology.tex
\section{Modeling}
\label{modeling}

\begin{figure*}
\begin{center}
\centering
\includegraphics[width=0.8\linewidth]{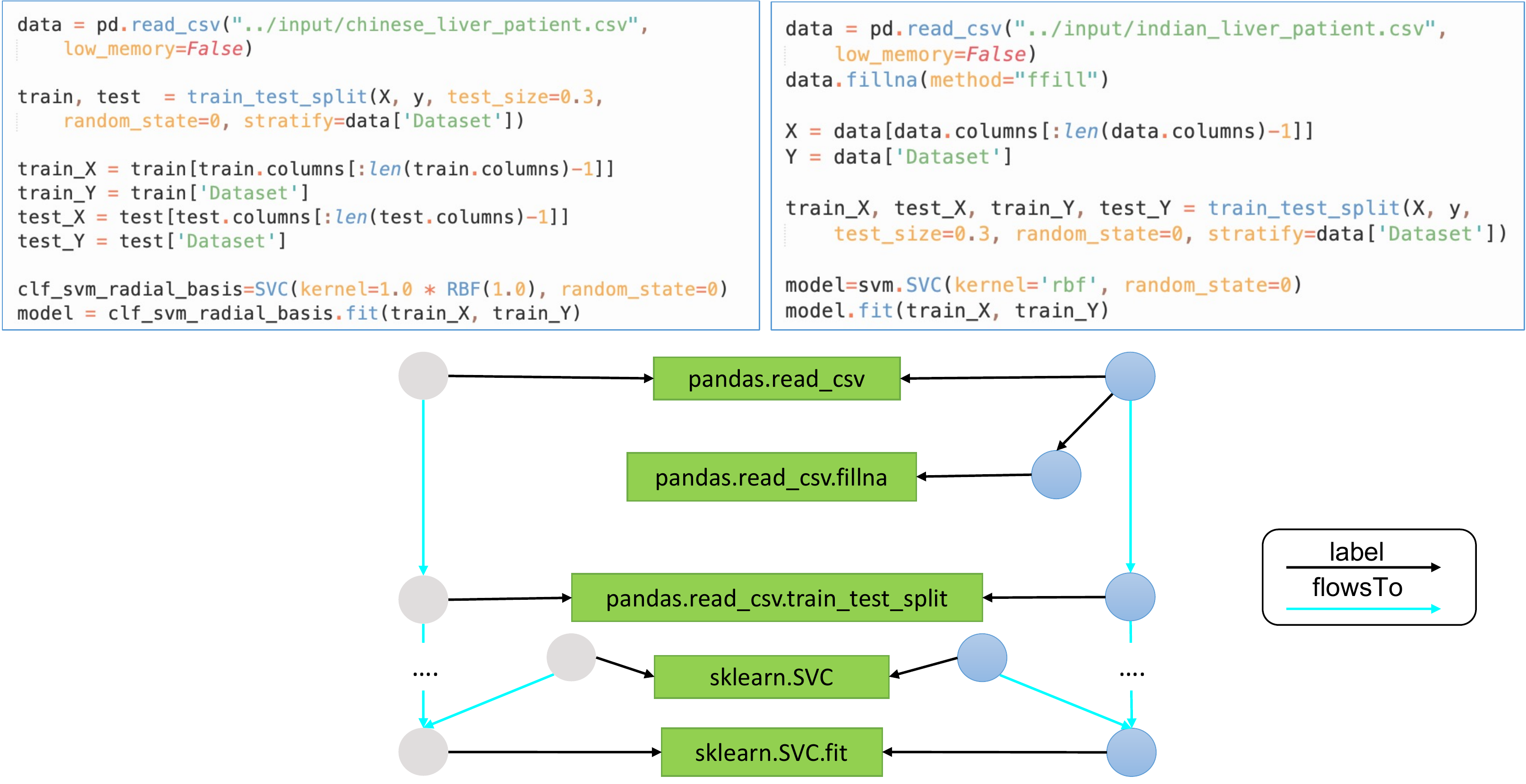}
\end{center}
\caption{Code analysis schema example for the code snippets in Figure \ref{fig:motivating_example} with gray and blue nodes corresponding to the code on the left and right, respectively.}
\label{fig:SA_example}
\end{figure*}

\sloppypar
%Figure~\ref{fig:schema} shows the schema adopted when the required output is RDF\footnote{The JSON output simply outputs nodes and edges of the graph, where each node has all the properties in the RDF version}.  When possible, we re-used properties from the Semanticscience Integrated Ontology (SIO) \cite{dumontier2014semanticscience}, and Schema.org, but we added a number of our own properties and classes, since we know of no single ontology that covers the concepts needed for modeling programs.  Our choice of Semanticscience Integrated Ontology (SIO) \cite{dumontier2014semanticscience}, and Schema.org was pragmatic; if a different modeling is needed, we do provide a JSON representation ADD CITATION HERE TO WEBSITE that is more skeletal and can be adapted to whatever ontology is required for an application.  We merely use this schema as a way to provide the user an intuitive understanding of what {\logo} captures.

%Classes and functions in code are translated into OWL Classes (see Figure~\ref{fig:schema} for example).  These OWL classes are also instances of OWL classes such as \texttt{graph4code:Class} and \texttt{graph4code:Function}.  Each class or function has a URI scheme based on the official python import path along with the \texttt{python:} prefix\footnote{http://purl.org/twc/graph4code/python/}.  
%Classes and functions in code are modeled as \texttt{g4c:Class} and \texttt{g4c:Function}\footnote{\texttt{g4c} is \url{http://purl.org/twc/graph4code/ontology/}}, and each has a URI based on its python import path along with the \texttt{python:} prefix\footnote{http://purl.org/twc/graph4code/python/}.

In modeling code graphs, a key design point is extensibility, which means loose coupling between multiple programs, and loose coupling to related bits of information about classes and functions.  To achieve this, we modeled each program as a separate graph, ensuring a unique node in the graph for every invocation of a call, or the read of any data structures such as lists, dictionaries, fields of an object, etc.  Each node is connected to its actual method or function name by an edge to a label node.  For instance, the function call for \texttt{pandas.read\_csv} in Figure~\ref{fig:motivating_example} would have a node that linked to its label which is \texttt{python:pandas.read\_csv}.  Two different invocations of the exact same function therefore only link to each other through their label, which would be \texttt{python:pandas.read\_csv}, as shown in Figure~\ref{fig:SA_example}.  However, since the return types of functions are often unknown in Python, this linkage between actual functions and the name for a particular invocation of the function is not always predictable.  For instance, in the right panel of Figure~\ref{fig:motivating_example}, the call \texttt{df.fillna} will reflect the analysis to that point as shown in Figure~\ref{fig:SA_example}, its label would be \texttt{pandas.read\_csv.fillna}, since the return value of a \texttt{pandas.read\_csv} call is unknown in dynamic languages such as Python and Javascript.  Modeling of code analysis is detailed in Section~\ref{sec:analysis}; Figure~\ref{fig:SA_example} demonstrates loose coupling across programs, and shows how code that looks different at a token level look similar at the data flow level.

%Each invocation of a function is an instance of \texttt{sio:SoftwareExecution}.  

% Add manchester notation listing for doctstrings examle for pandas.read_csv.

%Function invocations in code analysis link to their functions by RDFS label.  

\begin{figure}
\begin{center}
\centering
\includegraphics[width=1\linewidth]{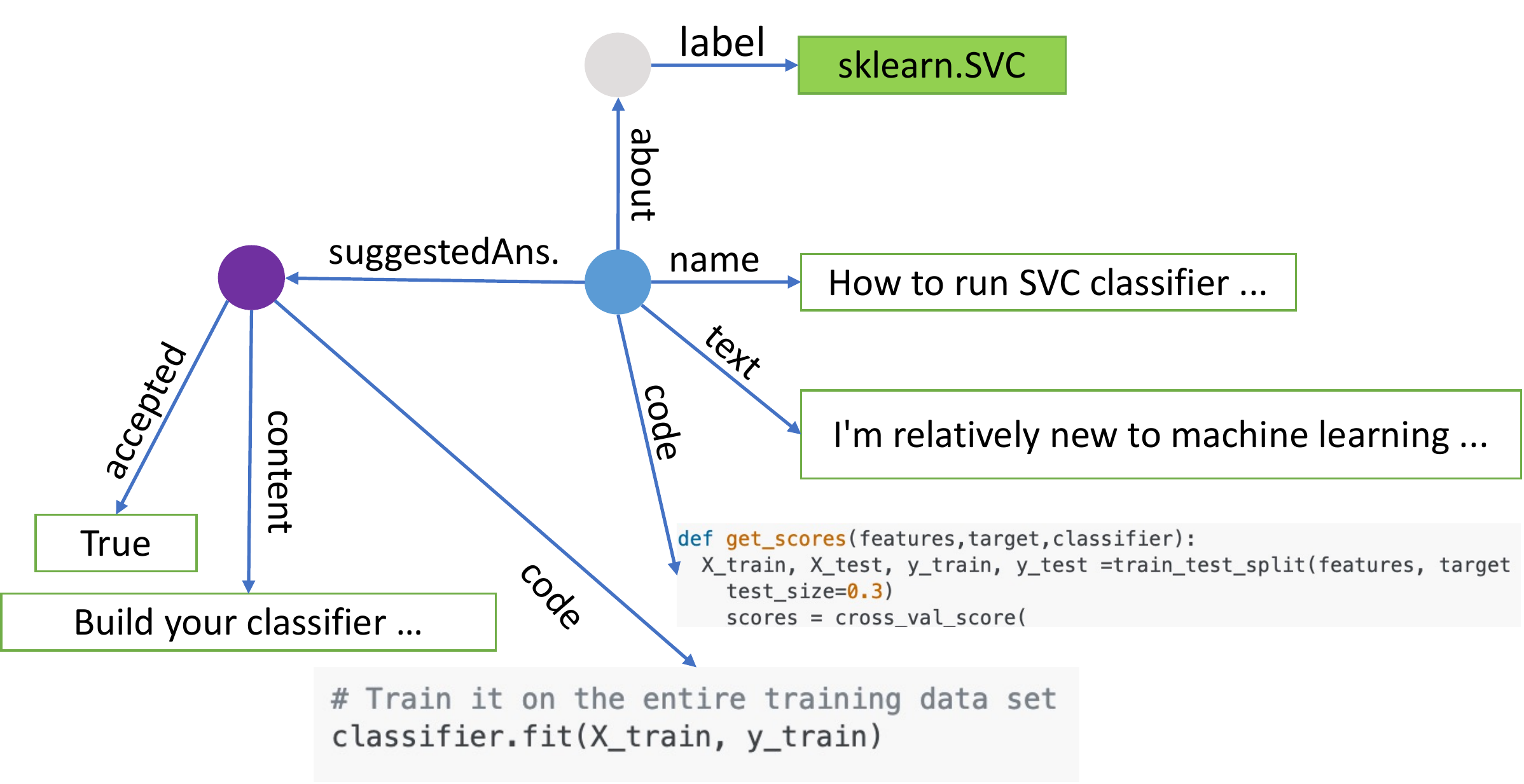}
\end{center}
\caption{StackOverflow Graph Example}
\label{fig:SO_schema}
\end{figure}

\begin{figure}
\begin{center}
\centering
\includegraphics[width=1\linewidth]{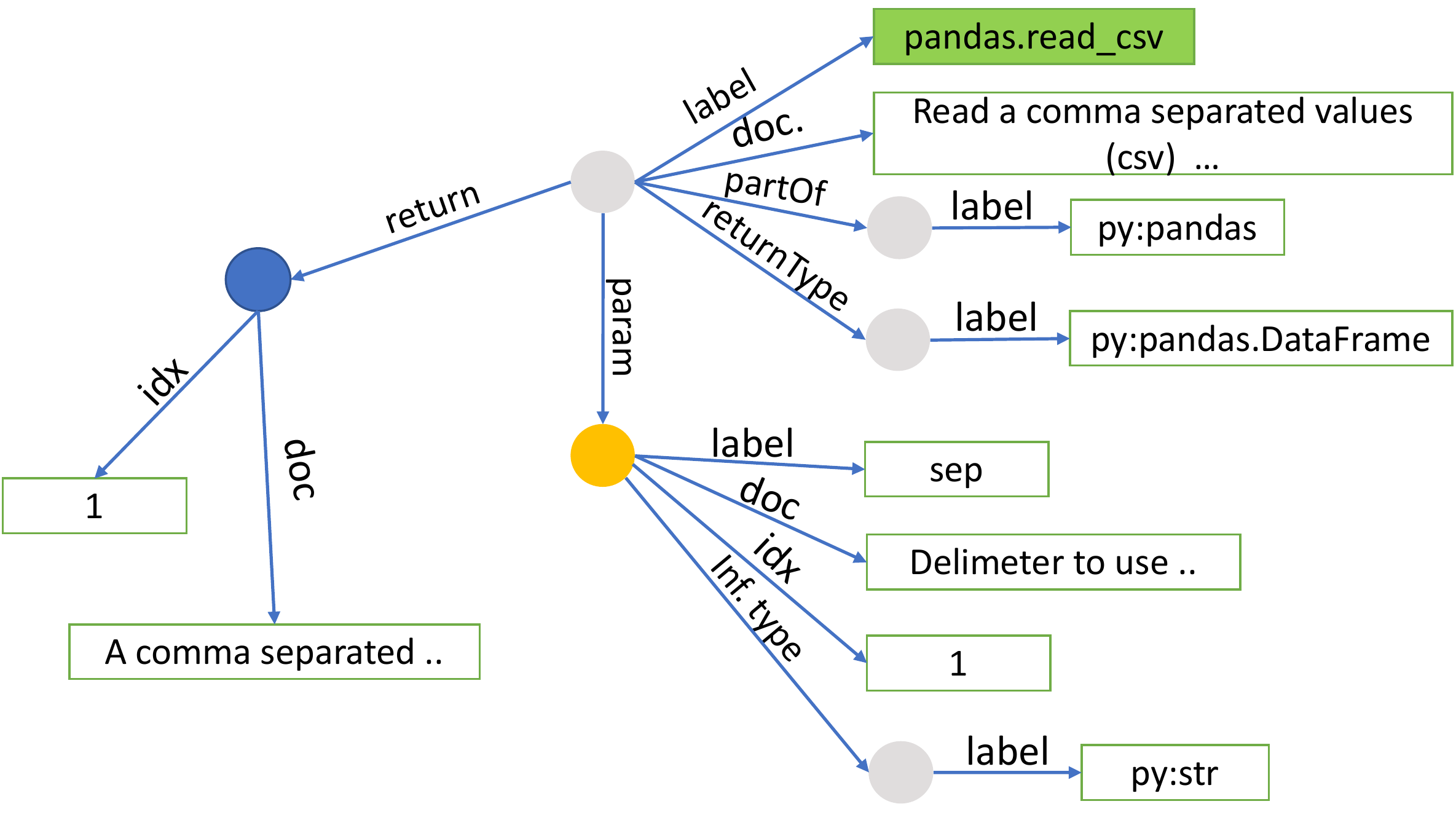}
\end{center}
\caption{Docstrings Graph Example}
\label{fig:docstrings_schema}
\end{figure}

%We model StackExchange questions and answers using properties and classes from Schema.org, while expressing the actual question text as \texttt{sioc:content}.  
%We chose Schema.org because it  models the social curation of questions and answers across the web.  
Similarly, each forum post has its own node, which describes the question, the answer (in terms of posts), any code mentioned in the posts (see Figure~\ref{fig:SO_schema}).  For documentation, once again the data is structured so a node representing documentation of a function would contain connections to the arguments and return types of that function (see Figure~\ref{fig:docstrings_schema}).  Loose coupling to program graphs is once again achieved by labels.  Posts with mentions of specific Python classes, modules, and functions (such as \texttt{SVC}) in the forum posts are linked to class and function labels in exactly the same manner as call invocations, i.e. with a link to a label node.  Similarly, documentation is connected to its class or function label.  Any new module's documentation, new program graphs, or new posts can be added to an existing graph, and the extraction process ensures that links to labels stitch new information in seamlessly.  For instance, a new forum post about \texttt{read\_csv} would link to the same label node as documentation of \texttt{read\_csv}.  The resulting graph allows the querying of usage patterns and natural language descriptions for python functions and classes directly by their fully qualified name (see Section~\ref{sec:kg}).

{\logo} produces output as RDF or JSON.  For the RDF version, to facilitate integration with other ontologies, we re-used properties from relevant existing ones such as the Semanticscience Integrated Ontology (SIO) \cite{dumontier2014semanticscience}, and Schema.org,  whenever possible. We, however, needed to add a number of our own properties and classes, since we found no single ontology that covers the concepts needed for modeling programs.  While we have found this representation useful, we do provide a JSON representation that can be adapted to whatever ontology is required for a given application if the current modeling is incompatible with a specific application's needs.  Details are available on the {\logo}'s  website.

%% file: sections/static_analysis.tex
\section{Mining Code Patterns}
\label{sec:analysis}

\subsection{Code Analysis}
\begin{figure}
\centering 
\includegraphics[width=0.95\linewidth]{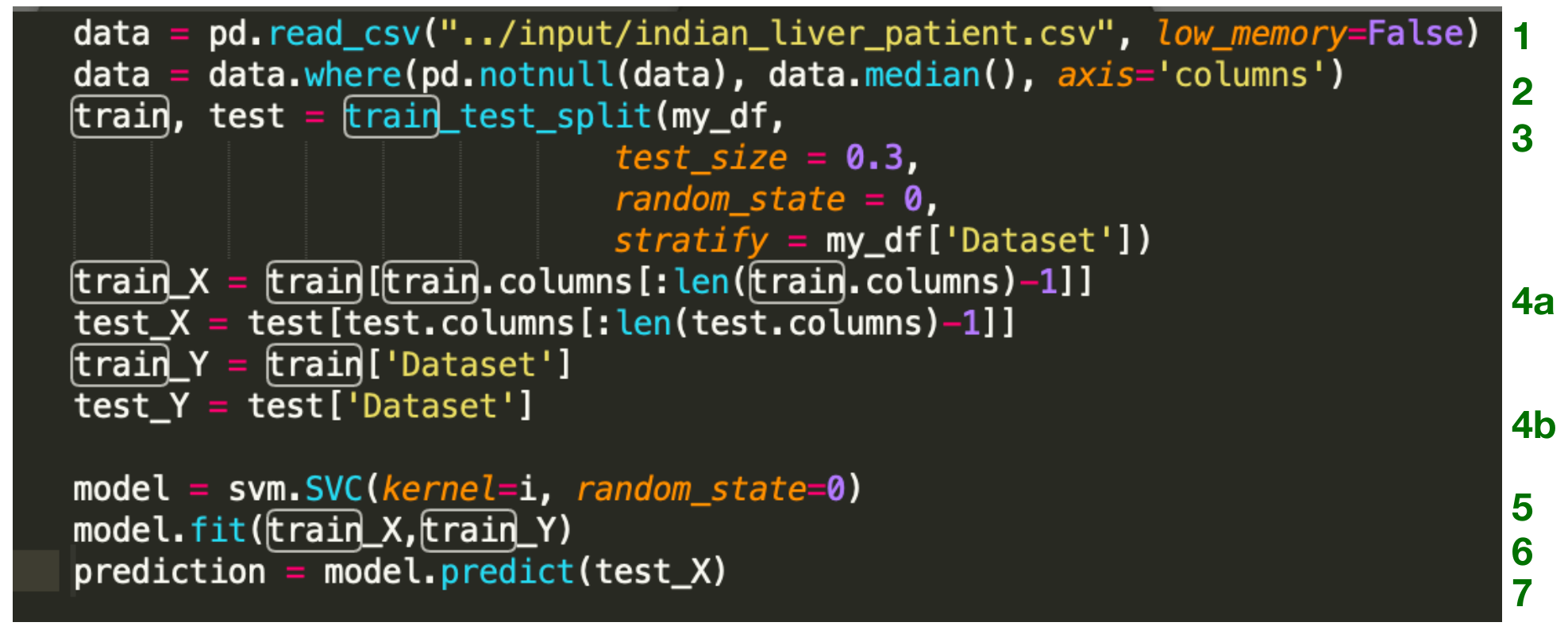}
\caption{Illustrative code example from GitHub}
\label{running_example}
\end{figure}

Although WALA is a well known library with extensive work (e.g. \cite{DBLP:conf/pldi/DolbySAR18,DBLP:conf/kbse/LeeDR16,DBLP:conf/ecoop/SridharanDCST12,10.1145/1985404.1985428,5601825}) on interprocedural data flow and control flow analysis for Java, Javascript and Python, we needed to extend the WALA analysis framework to make it work for millions of Python programs with many popular API uses. At the core is the issue of how to model the semantics of API functions.  There have been two basic options to it: (a) use the analysis framework to trace data and control flow into the code of the API function, (b) abstract it by using manually defined templates that makes the semantics of the call explicit.  The first approach is not feasible for Python because a vast portion of Python code wraps code written in a completely different language (e.g. C).  Moreover, given millions of lines of code in libraries behind the API calls, it is unlikely that the analysis can scale with adequate precision.  The second approach is tedious and can only work for a very small number of API calls.  In analyzing large numbers of programs with thousands of programs, this approach will not scale.  

To scale analysis to a large number of programs, we adopt a simple, common abstraction for all APIs; as with any static model of code~\cite{10.1145/2644805}. This abstraction is  neither sound nor complete, but it captures the essence of how APIs impact user code:
(a) We assume any call to an imported function simply returns a new object, with no side effects.  Of course this may not be true in practice, but it allows us to scale without delving into the analysis of libraries, which may be huge and, in languages like Python, are often written in C.  In other words, this is indeed a simplifying assumption that adds imprecision in the analysis, but there is no method to scale analysis to millions of programs without this assumption.
(b) We assume that any read of a field of an object created from an imported function returns itself.  Again, the field of an object is certainly not the same as the object itself, but it once again helps us scale analysis to large numbers of API types without having to capture the semantics of each of their fields, which may not be feasible in dynamic languages like Python.
(c) These new objects returned by the model may have methods called on them or fields read.  Such calls or reads on these objects are handled just like those on the imported API itself.  This allows calls like \texttt{read\_csv.where} to return new objects.

This extension to WALA's modeling framework is available only for Python at the moment, but a similar mechanism can be applied to other dynamic programming languages like Javascript.  Java has less need of this extension since it is a strongly typed language.

Figure~\ref{running_example} extends the code in our running example (Figure~\ref{fig:motivating_example}) to illustrate how we construct the analysis graph.  
In this example, a CSV file is first read using the Pandas library with a call to \texttt{pandas.read\_csv}, with the call being represented as \texttt{1} on the right of the program.  The object returned by the call has an unknown type, since most code in Python is not typed.  Some filtering is performed on this object by filling in the missing values with a median with a call to \texttt{where}, which is call \texttt{2}.  The object returned by \texttt{2} is then split into train and test with a call to \texttt{train\_test\_split}, which is call \texttt{3}.  Two subsets of the train data are created (\texttt{4}) which are then used as arguments to the \texttt{fit} call (\texttt{6}), after creating the \texttt{SVC} object in call (\texttt{5}).  The test data is similarly split into its X and Y components and used as arguments to the \texttt{predict} call (\texttt{7}).

\begin{figure}[htb]
\begin{center}
\centering
\includegraphics[width=0.8\linewidth]{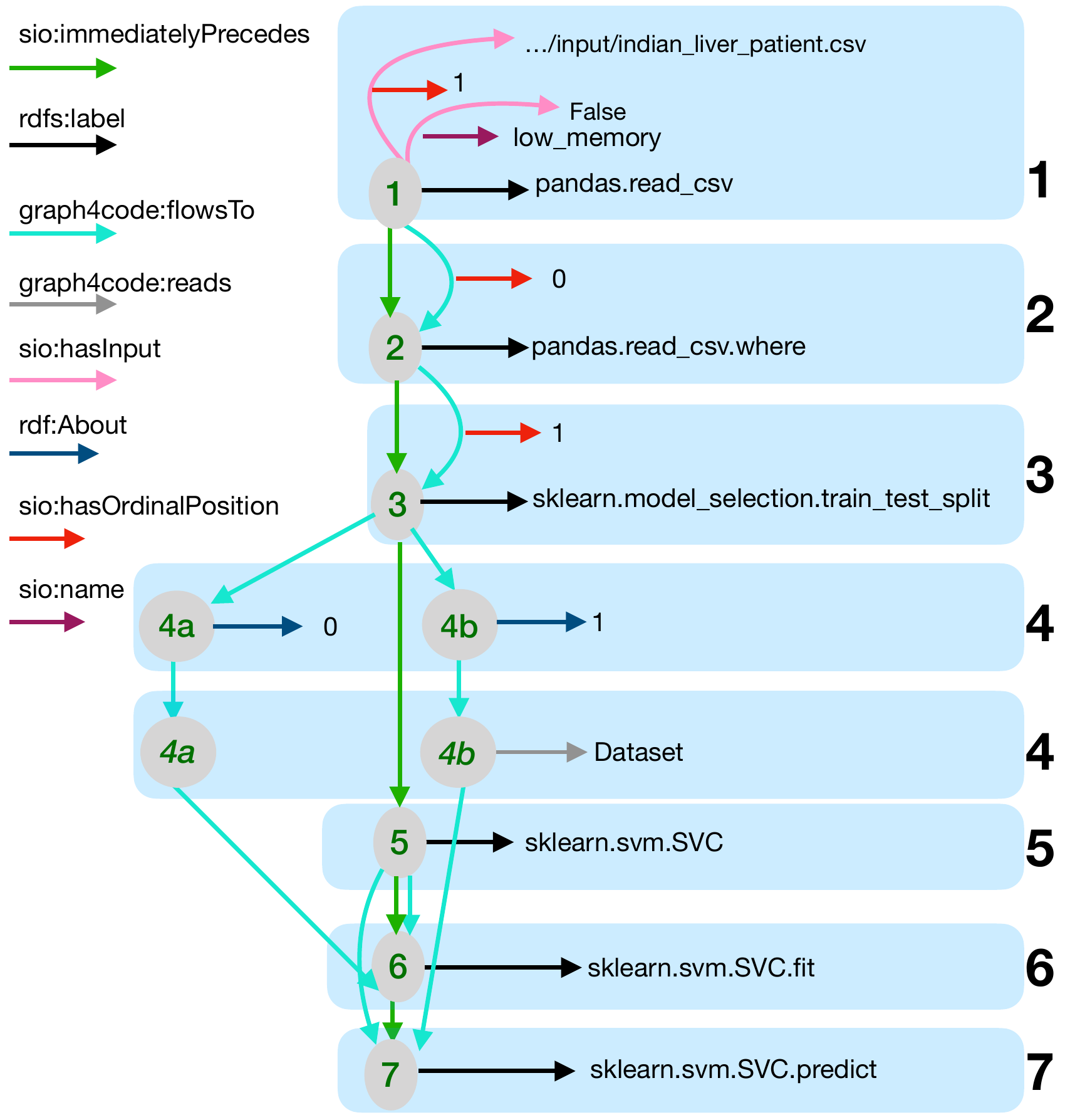}
\end{center}
\caption{Dataflow graph for the running example}
\label{fig:code_graph}
\end{figure}

Figure~\ref{fig:code_graph} illustrates the output of analysis, with control flow (depicted as green edges) and data flow (depicted as cyan edges) for the analyzed program.  The control flow edges are straightforward in this example and capture the order of calls, this is particularly useful when data flow is not explicit, such as when a \texttt{fit} call (labeled 6) must precede a \texttt{predict} call (labeled 7) for the \texttt{sklearn} library.  

We discuss the data flow shown in Figure~\ref{fig:code_graph} in more detail to provide an intuition of assumptions we made in our modeling to allow scalability to millions of programs.  This figure is a subset of the actual model, but we show all the key relations at least once.  This graph shows two key relations that capture the flow through the code:
\begin{description}
\item[flowsTo] (blue edges) captures the notion of data flow from one node to another, abstracting away details like argument numbers or names.  
%Many application queries can be expressed as transitive paths on \texttt{flowsTo}.  %We use SPARQL's property path operators extensively to accomplish this.
\item[immediatelyPrecedes] (green edges) captures code order.  %Queries such as \texttt{predict} calls not preceded by \texttt{fit} calls can be expressed this way.
\end{description}

In Figure~\ref{fig:code_graph}, nodes are labeled with numbers corresponding to the right hand side of Figure~\ref{fig:code_graph}, and the nodes are connected with edges that indicate control- and data-flow, as well as other properties.
Node \texttt{1} in Figure~\ref{fig:code_graph} corresponds to the execution of \texttt{read\_csv}.  Our graph captures arguments, which are the filename and the \texttt{low\_memory} option.  Since these are literals, the edges are in some sense backwards with respect to dataflow, since RDF does not allow edges from literals.

Node \texttt{2} corresponds to the \texttt{where} call, so it has both \texttt{immediatelyPrecedes} and \texttt{flowsTo} edges from node \texttt{1}, since it follows \texttt{1} in program order and gets data from it.  The \texttt{flowsTo} edge is annotated with a \texttt{hasOrdinalPosition} edge to \texttt{0}, indicating it is the receiver of the \texttt{where} call\footnote{In other words, its the object on which the call is made.}.  

Node \texttt{3} for \texttt{test\_train\_split} is similarly related to node \texttt{2}: it follows in program order, but data is passed as argument 1, so it is actually the first parameter to \texttt{test\_train\_split}.
The call to \texttt{test\_train\_split} returns a tuple, which is split into \texttt{train} and \texttt{test}.  This is captured in the first boxes labeled \texttt{4a,b}, which have labels denoting which value they receive.  

Then, in the code, each of \texttt{train} and \texttt{test} are split into their X and Y components for learning, which is shown in the italicized {\em{4a,b}}.  The \texttt{train} node is {\em{4a}} which is used for \texttt{fit}, and test {\em{4b}} node shows the read from \texttt{Dataset}.  The {\em{4b}} nodes flow to the \texttt{predict} call; the X field is a slice and so does not have a specific field.  Note that this example does not have dataflow directly from \texttt{fit} to \texttt{predict}, but this graph also captures the ordering constraint between them as shown in Figure~\ref{fig:code_graph}.
Nodes \texttt{5,6,7} similarly show control- and data-flow of the use of \texttt{SVC}.

\begin{figure}[htb]
\begin{center}
\centering
\includegraphics[width=1.0\linewidth]{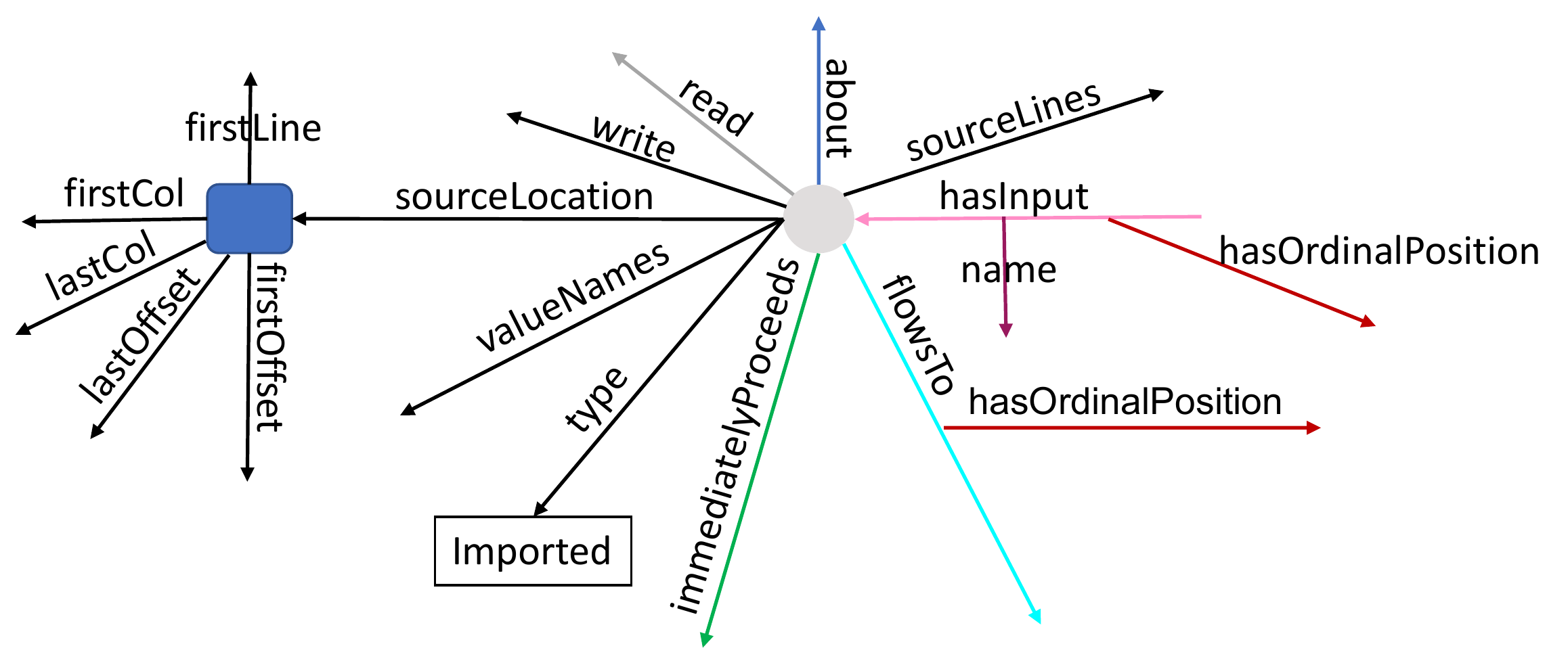}
\end{center}
\caption{Code Analysis Schema}
\label{fig:code_schema}
\end{figure}

Figure~\ref{fig:code_schema} shows the schema of each node in the analysis graph.  Apart from edges described in the running example, each node comes with a definition of corresponding \texttt{sourceLocation} in the original code, its \texttt{value\_names} which describes the local variables that a corresponding call got written to, in case an application needs it, whether the node reflects an import call, if the node \texttt{reads} from or \texttt{writes} to some object returned by a specific call.

%Each program is analyzed into a separate graph, with connections between different programs being brought about by common calls (e.g., \texttt{sklearn.svm.SVC.fit}).  Note that this type of modeling  accommodates the addition of more graphs easily as more code gets available for analysis.  

\subsection{Extraction of Python Files from GitHub }
To test the scalability of {\logo}, we ran the toolkit on 1.38 million code files from GitHub.  To extract this dataset, we ran a SQL query on Google BigQuery dataset\footnote{\url{https://github.com/wala/graph4code/blob/master/extraction_queries/bigquery.sql}}.  The query looks for all Python files and Python notebooks from repositories that had at least two watch events associated with it, and excludes large files.  Duplicate files were eliminated, where a duplicate was defined as having the same MD5 hash as another file in the dataset.  All the files were then analyzed to produce data flow and control flow graphs for each script.  For each script to provide maximal coverage of the code, we added entry points for each defined function in each script, as well as the main body.

%% file: sections/extraction_linking.tex
\section{Linking Code to Documentation and Forum Posts}
\label{sec:extraction}

%%%%%%%%%%%%%%%%%%%%%%%%%%%%%%%%%%%%%%%%%%%%%%%%%%%%%%%%%%%
%%%%%%%%%%%%%%%%%%%%%%%%%%%%%%%%%%%%%%%%%%%%%%%%%%%%%%%%%%%
%%%%%%%%%%%%%%%%%%%%%%%%%%%%%%%%%%%%%%%%%%%%%%%%%%%%%%%%%%%
%%%%%%%%%%%%%%%%%%%%%%%%%%%%%%%%%%%%%%%%%%%%%%%%%%%%%%%%%%%

\subsection{Extracting Documentation into the Graph}
\label{documentation_docstrings}

To generate documentation for all functions and classes used in the 1.3 million files, we first processed from the analysis step above all the import statements to gather popular libraries (libraries with more than 1000 imports across  files).  506 such libraries were identified, of which many were part of Python language itself.  %For each of these libraries, we tried to resolve their location on GitHub to get not only the documentation embedded in its code, but also associated usage documentation which tends to be in the form of markdown or restructured text files.  We found 403 repositories using web searches on GitHub for the names of modules. 

% \sloppypar
%The first source of documentation we collected is from the documentation embedded in these code files.  
%Of a total of 257,081 files with the extension of `.py' across the 403, plus all the Python2 and Python3 modules, 167,191 had functions, classes or methods that produced some documentation in embedded in code.  10,841 files were empty (in Python, modules usually have an empty file to mark the module), 444 files failed to parse, and 51,900 files were dropped from analysis because they were likely tests, and 37,546 did not produce any parsed functions or classes in the AST despite not being empty (this is possible in a scripting language like Python).  
% We first filtered all code files that are empty, having parsing issues or are just testing modules. 
%However, documentation in the source is insufficient because Python libraries often depend on code written in other languages.  As an example, the {\tt tensorflow.placeholder} function is defined in C, and has no stub that allows the extraction of its documentation.  Therefore, to gather additional documentation 

For these libraries, we used Python introspection techniques to created a virtual environment, install the library, and used {\tt inspect} to gather the documentation. Clearly this step is language specific - the toolkit code currently only has code to perform extraction from Python code.  However, we note that extraction of documentation for code is supported in multiple languages, and so any extension to a new language requires language specific code for this part.
%We were successful in gathering documentation only for 432 modules because of software dependency issues; dependence on OS libraries or Python versions prevented successful installs in other cases\footnote{In general, we do not handle versioning well, and this is clearly an area for future work.}.  
%The introspection code ran on Python 3.7.  This step generated an additional 17,587 files with documentation about classes, methods and functions, for an additional 55,306 pieces of code documentation. 

This step yielded 6.3M pieces of documentation for functions, classes and methods in 2300+ modules (introspection of each module brought in its dependencies). The extracted  documentation is added to our code knowledge graph where, for each class or function, we store its documentation string, base classes, parameter names and types, return types and so on. An example of such extracted information is shown in Figure \ref{fig:docstrings_schema}. The full RDF version of this 6.3M edges graph is available in {\logo}'s website. 

\subsection{Extraction of StackOverflow and StackExchange Posts}
\label{documentation_stackoverflow}

User forums such as Stackoverflow\footnote{\url{https://stackoverflow.com/}} provide a lot of information in the form of questions and answers about code. Moreover, user votes on questions and answers can indicate the value of the question and the correctness of the answer. While Stackoverflow is geared towards programming, StackExchange\footnote{\url{https://data.stackexchange.com/}} provides a network of sites around many other topics such as data science, statistics, mathematics, and artificial intelligence. 

To further enrich our knowledge graph with natural language posts about code and other documentation, we linked each function, class and method to its relevant posts in Stackoverflow and StackExchange. In particular, we extracted 45M posts (question and answers) from StackOverflow and 2.7M posts from StackExchange in Statistical Analysis, Artificial Intelligence, Mathematics and Data Science forums. Each question is linked to all its answers and to all its metadata information like tags, votes, comments, codes, etc.

We then built an elastic search index across these sources, where each document is a single question with all its answers. The documents were indexed using a custom analyzer tailored for natural language as well as code snippets. 
Then, for each function, class and method, we perform a multi-match search\footnote{\url{https://github.com/wala/graph4code/blob/master/extraction_queries/elastic_search.q}} over this index to retrieve the most relevant posts (a  limit of 5K matches per query is imposed) and link it to the corresponding node in the knowledge graph. 
This step generalizes nicely across languages since all we do here is gear the analyzer and search queries to the idiosyncracies of code. Figure \ref{fig:SO_schema} shows an example of how these information is structured in {\logo} and how it links to docstrings and 
code analysis through label (green) nodes. The full RDF version of this graph can be found in \url{https://wala.github.io/graph4code/}. 
% A sample graph depicting the schema of extracted forum posts is available in \url{https://wala.github.io/graph4code/}.

With this extraction and linking mechanism, one can leverage the plethora of links we have from posts to different classes for code recommendation. For example, the post in Figure  \ref{fig:motivating_example} discusses the usage of \textit{sklearn.SVC} class for a machine learning problem. Similarly, the post \url{https://stackoverflow.com/questions/33840569/} talks about \textit{sklearn.SVC} and how it suffers from memory issues when handling larger datasets. It also mentions \textit{linear\_model.SGDClassifier} as a solution to this problem. With {\logo}, the two  posts will have a link to \textit{sklearn.SVC} while the second post will have an extra link to \textit{linear\_model.SGDClassifier} which one could use for code recommendation. 

\subsection{Extracting Class Hierarchies}
\label{documentation_classhierarchy}

\sloppypar
As in the case of extracting documentation embedded in code, extraction of class hierarchies was based on Python introspection of the 2300+ modules. For example, the below triples list some of the subclasses of \texttt{BaseSVC}:

{
% \scriptsize
\begin{verbatim}
py:sklearn.svm.SVC rdfs:subClassOf 
                    py:sklearn.svm._base.BaseSVC . 
py:sklearn.svm.NuSVC rdfs:subClassOf 
                    py:sklearn.svm._base.BaseSVC .
\end{verbatim}
}
This step again like the extraction of documentation need some customization for each new language that \logo\   can support over time.

%% file: sections/stats.tex
\section{Extracted Code Knowledge Graph: Properties and Uses}
\label{sec:kg}
%As discussed earlier, a large literature exists on using various code abstractions such as ASTs, text, or even output of program analysis artifacts for all sorts of applications such as code refactoring, code search, code de-duplication detection, debugging, enforcement of best practices etc.  The popularity of WALA\footnote{\url{https://github.com/wala/WALA}}, which we leverage   for program analysis, attests to the fact that numerous applications exist for this type of representation of code.  To our knowledge, {\logo} is the first toolkit geared to build code knowledge graphs over a large repository of programs and systematically link it to documentation and forum posts related to code.  We believe that by doing so, we will enable a new class of applications that  combine code semantics as expressed by program flow along with natural language descriptions of code.

\label{graph_properties}
%dlm could use an intro before the first subsection.  this section is one of the important contributions and one that people might use - bring that out

\subsection{Graph Statistics}
\label{sec:graph_stats}

\begingroup
\setlength{\tabcolsep}{3pt}
\renewcommand{\arraystretch}{0.5}
\begin{table}[t]
\centering
\small
\begin{tabular}{lccc}
\toprule
                & Functions(K) & Classes(K) & Methods(K)\\
\midrule
Docstrings      &   278        &  257       &   5,809     \\
\midrule
Web Forums Links  &    106      &    88     &  742    \\ 
Static Analysis Links&   4,230        &    2,132     &    959   \\
\toprule
\end{tabular}
\caption{Number of functions, classes and methods in docstrings and the links connected to them from user forums and static analysis. This results in a knowledge graph with 2.09 billion edges in total}
\label{link_analysis}
\end{table}
\endgroup

% Quad counts:docstrings: 722,096,354
% stackoverflow/stackexchange: 89,540,211
% turtle analysis: 1,171,325,098

Table~\ref{link_analysis} shows the number of unique methods, classes, and functions in docstrings for our sample extraction (embedded documentation in code). These correspond to all documentation pieces we found embedded in the code files or obtained through introspection. Overall, we extracted documentation for 278K functions, 257K classes and 5.8M methods. 
Table \ref{link_analysis} also shows the number of links made from other sources to docstrings documentation. Static analysis of the 1.3M code files created a total of 7.3M links (4.2M functions, 2.1M class and 959K methods). We also created links to web forums in Stackoverflow and StackExchange: {\logo} has currently 106K, 88K and 742K links from web forums to functions, classes and methods, respectively. This results in a knowledge graph with a total of 2.09 billion edges; 75M triples from docstrings, 246M from web forums and 1.77 billion from static analysis.

\subsection{Querying RDF output of {\logo}}
This section shows basic queries for retrieving information from RDF generated by {\logo} again purely for illustrative purposes on what is represented in the code graph. 

The first query returns the documentation of a class or function, in this case \texttt{pandas.read\_csv}. It also returns parameter and return types, when known. One can expand these parameters (\texttt{?param}) further to get their labels, documentation, inferred types, and check if they are optional.

{
% \scriptsize 
\begin{verbatim}
select ?doc ?param ?return where {
graph <http://purl.org/twc/graph4code/docstrings> 
    {
      ?s  rdfs:label            "pandas.read_csv" ;
          skos:definition       ?doc .
      optional { ?s graph4code:param   ?param . }
      optional { ?s graph4code:return  ?return . }
    }
}
\end{verbatim}    
}
% {\small 
% \begin{verbatim}
% select ?doc ?param ?return where {
%   ?s  rdfs:label "pandas.read_csv" ;
%       skos:definition ?doc ;
%       rdfs:subClassOf [
%         owl:onProperty sio:SIO_000230; # has input
%         owl:someValuesFrom ?s
%       ] ;
%       rdfs:subClassOf [
%         owl:onProperty sio:SIO_000229 # has output
%         owl:someValuesFrom ?return
%       ] . 
%     }
% } 
% \end{verbatim}    
% }
In addition to the documentation of \texttt{pandas.read\_csv}, we can also get the forum posts that mention this function by appending the following to the query above. This will return all questions in StackOverflow and StackExchange forums about \texttt{pandas.read\_csv} along with its answers. 

{
% \scriptsize 
\begin{verbatim}
  graph ?g2 {
        ?ques   schema:about            ?s ;
                schema:name             ?q_title ;
                schema:suggestedAnswer  ?a .
        ?a sioc:content ?answer.
    }
\end{verbatim}    
}

Another use of a code knowledge graph produced by {\logo} is to understand how people use functions such as \texttt{pandas.read\_csv}. In particular, the query below shows when \texttt{pandas.read\_csv} is used, what are the \texttt{fit} functions that are typically applied on its output. 
%We have used the results of a similar query to drive code recommendations in an IDE.
%TODO: is this redundant? correct? 

{
% \scriptsize 
\begin{verbatim}
select distinct ?label where {
  graph ?g {
    ?read   rdfs:label          "pandas.read_csv" .
    ?fit    schema:about        "fit" .
    ?read   graph4code:flowsTo+ ?fit .
    ?fit    rdfs:label          ?label .
  }
}
\end{verbatim}    
}

\subsection{Assessing the Quality of the Sample Extraction}
To assess the validity of the control flow and data flow graphs produced by {\logo} with the specific additions we made to WALA to analyze a large number of Python programs, we analyzed the ASTs of 441 Python programs.   The graphs produced by WALA abstract away variables, and track the flow of objects through function calls. Therefore, we wanted to assess to what degree WALA captures function calls present in the program code in its analysis graphs.  For the ASTs, we recorded all the locations of \textit{call} nodes in the program code, because they record function calls in the program.  For these call nodes, we tracked whether the specific call was covered in our WALA graphs, at the same source location.  Across 441 programs, WALA generated dataflow and control flow graphs that covered an average of 86\% of calls (standard deviation was 13\%).  Given that static analysis is neither sound nor complete~\cite{10.1145/2644805}, the graphs seem to capture most of the calls.

% Documentations such as docstrings and  

We also wanted to assess the quality of the links {\logo} produced for forum posts\footnote{There was less need to assess the quality of documentation since that was generated by Python introspection techniques}. To do so, we extracted a sample of 100 links between classes, functions and methods and their associated forum posts.  We then asked two human annotators to judge the quality of these links as being relevant or not. A relevant link would mandate that the post is discussing the same exact module, class, and function, otherwise it is considered irrelevant. The two annotators had a 100\% agreement and both measured a 79\% linking accuracy. Analyzing the irrelevant links showed that it typically happens across languages where the same class appears in multiple languages or in generic classes or builtin types. Examples of such cases include like \textit{Formatter} and \textit{AssertionError} classes in Python and Java respectively.

%% file: sections/related.tex
\section{Related Work}

To our knowledge, there is no comprehensive knowledge graph for code that integrates semantic analysis of code along with textual artifacts about code.  Here we review related work around issues of how code has been typically represented in the literature, what sorts of datasets have been available for code, and ontologies or semantic models for code.

\subsection{Code Representation}
A vast majority of work in the literature has used either tokens or abstract syntax trees as input representations of code (see \cite{Allamanis:2018:SML:3236632.3212695} for a comprehensive survey on the topic).  
When these input code representations are used for a specific application, the target is usually a distributed representation of code (see again \cite{Allamanis:2018:SML:3236632.3212695} for a breakdown of prior work), with a few that build various types of probabilistic graphical models from code.  
Few works have used data and control flow based representations of code as input to drive various applications.  As an example, \cite{DBLP:conf/icpp/HsiaoCN14} used a program dependence graph to detect code duplication on Javascript programs, but the dependence is computed in an intra-procedural manner.  Similarly, \cite{DBLP:conf/iclr/AllamanisBK18} augments an AST based representation of code along with local data flow and control flow edges to predict variable names or find the misuse of variables in code.  \cite{DBLP:journals/corr/abs-1811-01824} combines token based representations of code with edges based on object uses, and AST nodes to predict the documentation of a method. 
%\cite{Chae:2017:AGF:3152284.3133925} uses C-reduce, which takes a C file, and a specific property of interest (e.g. an assertion about values of a particular variable in a program), and builds an abstract data flow over this reduced C program to understand what sort of data flow patterns require expensive static analysis techniques such as flow sensitivity.
\cite{Bruch:2009:LEI:1595696.1595728, Proksch:2015:ICC:2852270.2744200} includes partial object use information from WALA for code completion tasks, but the primary abstraction in that work is (a) a vector representation of APIs for Java SWT, that they used in machine learning algorithms such as best matching neighbors to find the next best API for completion \cite{Bruch:2009:LEI:1595696.1595728}, or (b) as a Bayesian network which reflects the likelihood of a specific method call given the other method calls that have been observed \cite{Proksch:2015:ICC:2852270.2744200}. \cite{Nguyen:2015:GSL:2818754.2818858, Nguyen:2009:GMM:1595696.1595767} employs a mostly intra-procedural analysis  to mine a large number of graphs augmented with control and data flow, for the purposes of code completion for Java API calls.  This work is interesting because, like us, \cite{Nguyen:2015:GSL:2818754.2818858} it creates a large program graph database which models dependencies between parent and child graphs, from which a Bayesian model is constructed to predict the next set of API calls based on the current one.  

Our work can be distinguished from prior work in this area in two key ways: (a) our work targets inter-procedural data and control flow, in the presence of first class functions and no typing, to create a more comprehensive representation of code, and (b) we use this representation to drive the construction of a \textit{multi-purpose} knowledge graph of code that is connected to its textual artifacts. 

% % Moved from introduction, might need to integrate better.
% In other contexts, nothing precludes applications from using the surface representations of programs by themselves for a specific task.  
% However, we note that prior work in many applications such as finding security bugs in code (\cite{DBLP:journals/corr/abs-1807-06756}), predicting variable names, method names, or types (\cite{DBLP:journals/corr/abs-1803-09544}), code summarization (\cite{DBLP:journals/corr/abs-1708-01837}), clone detection (\cite{White:2016:DLC:2970276.2970326}), predicting variable misuse, variable naming, deobfuscation, and method naming (\cite{DBLP:journals/corr/abs-1711-00740}, \cite{Bichsel:2016:SDA:2976749.2978422}, \cite{DBLP:journals/corr/abs-1808-01400}), dataflow and control flow have been selectively added to surface representations of programs to gain a performance advantage.  
% Our approach provides this capability for millions of programs in a more general way, using state-of-the-art techniques to follow object use across procedure calls, and to model heap usage. 

\subsection{Code Datasets}
% To the best of our knowledge, there is no work that tries to build a general knowledge graph for code.  can directly benefit from our resource graph such as code summarization (https://arxiv.org/abs/2005.00653, https://github.com/tech-srl/code2vec), recent code search (CodeSearchNet) by Facebook (https://arxiv.org/abs/1909.09436) and models such as CodeBERT ( https://arxiv.org/abs/2002.08155).   There are also existing datasets for each of these tasks, which we will also highlight, but they tend to be specific to code- and task-specific.
Several research efforts started recently to focus on using machine learning for code summarization~\cite{ahmad2020transformer,iyer2016summarizing,alon2019code2vec}, code search~\cite{husain2019codesearchnet} and models such as CodeBERT~\cite{feng2020codebert}. 
The datasets used by these approaches tend to be code- and task-specific. To the best of our knowledge, there is no work that tries to build a general knowledge graph for code and we believe that these approaches can directly benefit from {\logo}. We, however, leave this for future work.

\subsection{Semantic Models of Code}

SemanGit \cite{kubitza2019semangit} is a linked data version based on Github activities. Unlike {\logo}, SemanGit focus on modeling user activities on Github and not on understanding the code itself as in {\logo}. 
CodeOntology \cite{atzeni2017codeontology} in an ontology designed for modeling code written in Java. The ontology is similar to ours when it comes to modeling relationships among classes and methods. A crucial difference, however, is how the code itself is parsed and hence how it gets modeled. CodeOntology's parser relies on Abstract Syntax Trees (AST) for understanding the semantics of the code while {\logo} represents programs as data flow and control flow which is crucial because programs that behave similarly can look arbitrarily different at a token or AST level due to syntactic structure or choices of variable names.  
%Furthermore, the referenced patent uses ASTs, which makes it dependent on, if not Java, then at least Java-like languages with similar ASTs.  Our dataflow-based approach generalizes over more languages.
\cite{jiomekong2019extracting} proposed an approach for learning ontology from Java code using Hidden Markov Models. Unlike this approach, {\logo} relies on a standard static analysis library (WALA) for code understanding which is then modeled using our ontology proposed earlier. \cite{cao2019unsupervised} aims to construct knowledge graph of scientific concepts expressed in text books to link it to code artifacts such as function or variable names. In their work, text sentences are converted into triples and linked to source code based on token matches. This is different from {\logo}'s approach where the focus is on modeling control and data flow of code and then linking classes and methods to their documentation. Augmenting {\logo} with an approach specified in \cite{cao2019unsupervised} is an interesting idea for future work.

% Kite - AI Coding Assistant for Python and JavaScript: https://www.kite.com/